\documentclass[sigconf]{acmart}

\AtBeginDocument{%
  }

\setcopyright{acmcopyright}
\copyrightyear{2024}
\acmYear{2024}
\acmDOI{XXXXXXX.XXXXXXX}

\acmConference[ICBBT'24]{Make sure to enter the correct
  conference title from your rights confirmation emai}{May 24--26,
  2024}{Chongqing, China}
\acmPrice{15.00}
\acmISBN{978-1-4503-XXXX-X/18/06}

\begin{document}

\title{Assessing Patient Eligibility for Inspire Therapy through Machine Learning and Deep Learning Models}

\author{Mohsena Chowdhury}
\affiliation{%
  \institution{Toronto Metropolitan University}
  \city{Toronto}
  \state{ON}
  \country{Canada}
}
\email{mohsena.chowdhury@torontomu.ca }

\author{Tejas Vyas}
\affiliation{%
  \institution{Toronto Metropolitan University}
  \city{Toronto}
  \state{ON}
  \country{Canada}
}
\email{tejas.vyas@torontomu.ca}

\author{Rahul Alapati}
\affiliation{%
  \institution{University of
Kansas Medical Center}
  \city{Kansas City}
  \state{KS}
  \country{USA}
}
\email{ralapati@kumc.edu}

\author{Andrés M Bur}
\affiliation{%
  \institution{University of
Kansas Medical Center}
  \city{Kansas City}
  \state{KS}
  \country{USA}}
\email{abur@kumc.edu }

\author{Guanghui Wang}
\affiliation{%
  \institution{Toronto Metropolitan University}
  \city{Toronto}
  \state{ON}
  \country{Canada}}
  \email{wangcs@torontomu.ca}

\renewcommand{\shortauthors}{Chowdhury et al.}

\begin{abstract}
Inspire therapy is an FDA-approved internal neurostimulation treatment for obstructive sleep apnea. However, not all patients respond to this therapy, posing a challenge even for experienced otolaryngologists to determine candidacy. This paper makes the first attempt to leverage both machine learning and deep learning techniques in discerning patient responsiveness to Inspire therapy using medical data and videos captured through Drug-Induced Sleep Endoscopy (DISE), an essential procedure for Inspire therapy. To achieve this, we gathered and annotated three datasets from 127 patients. Two of these datasets comprise endoscopic videos focused on the Base of the Tongue and Velopharynx. The third dataset composes the patient's clinical information. By utilizing these datasets, we benchmarked and compared the performance of six deep learning models and five classical machine learning algorithms. The results demonstrate the potential of employing machine learning and deep learning techniques to determine a patient's eligibility for Inspire therapy, paving the way for future advancements in this field.
\end{abstract}



\maketitle

\section{Introduction}
Obstructive sleep apnea is a common sleep disorder that impacts millions of people worldwide. It is characterized by repetitive episodes of partial or complete obstruction of the upper airway during sleep, leading to fragmented sleep and decreased oxygen levels in the blood. Treatment for obstructive sleep apnea aims to reduce daytime sleepiness and the morbidity and mortality associated with increased risks of ischemic heart disease, cardiac arrhythmias, hypertension, and other vascular complications \cite{gasparini2021functional}\cite{levy2015obstructive}.

In previous studies, AI-based technologies have demonstrated great potential in the diagnosis and treatment of patients with obstructive sleep apnea. By utilizing AI in sleep medicine, clinicians can enhance their ability to accurately diagnose and tailor treatment plans for individual patients \cite{molnar2022predictive}\cite{su2023development}. AI technologies can analyze sleep patterns and identify specific markers of obstructive sleep apnea, allowing for more efficient and accurate diagnoses. They can also assist in monitoring the effectiveness and adherence of treatment, making adjustments as needed to optimize patient outcomes \cite{huang2022predicting}\cite{molnar2022predictive}\cite{van2021comparison}. Additionally, AI can aid in identifying patients who may not respond well to traditional treatment methods, such as continuous positive airway pressure or mandibular advancement devices, guiding clinicians in considering alternative therapies like positional therapy \cite{brennan2023role}.

Most prior AI-based studies that utilized machine learning and predictive analytics, can analyze large amounts of clinical data to identify patterns and markers of obstructive sleep apnea \cite{huang2022predicting}\cite{molnar2022predictive}. In recent years, the classification of snores produced in different airway states used ML-based guided treatment for OSA before Drug-Induced Sleep Endoscopy (DISE) \cite{huang2022predicting}\cite{liu2022automatic}. There is very limited use of deep learning-based approaches to identify obstructive sleep apnea from DISE videos for identifying the airway collapse patterns and location \cite{hanif2021upper} and severity scores \cite{hanif2023automatic}.

In this study, we employ endoscopic images obtained from the DISE procedure, where patients are sedated to simulate sleep and the upper airway is evaluated to determine eligibility for surgical treatment for obstructive sleep apnea (OSA). This research aims to predict which patients will respond to Inspire, a surgical implant designed to stimulate tongue movement. The prediction outcome seeks to categorize patients as Responders or Non-responders based on the analysis of endoscopy images. This study is the first endeavor to predict a patient’s response emphasis on the base of tongue (BOT) or velopharynx (VP) throat region images. We implemented and compared the performance with five machine learning algorithms: Decision Tree \cite{ke2017lightgbm}, Gradient Boosting \cite{ke2017lightgbm}, k-nearest Neighbors \cite{peterson2009k}, Logistic Regression \cite{sperandei2014understanding}, and Random Forest \cite{biau2016random}, as well as six deep learning models: VGG-16 \cite{simonyan2014very}, ResNet-50 \cite{jian2016deep}, ResNet-101 \cite{jian2016deep}, EfficientNet-B0 \cite{tan2019efficientnet}, DenseNet-121 \cite{huang2017densely}, DenseNet-169 \cite{huang2017densely}.

The major contributions of this paper are as follows:
\begin{itemize}
\item We conducted the first study with the objective of predicting the patient response to assess eligibility for Inspire therapy using endoscopy images from Drug-Induced Sleep Endoscopy videos.
\item We generated and annotated three datasets from a cohort of 127 patients, totaling 24,750 image frames. The dataset encompasses 88 cases of responders and 39 cases of non-responders.
\item We implemented and benchmarked the performance of five machine learning algorithms and six deep neural network models utilizing the generated datasets.
\end{itemize}

\section{Background}

Obstructive Sleep Apnea (OSA) is a widespread health disorder marked by recurrent instances of upper airway collapse during sleep. This condition leads to disrupted sleep patterns and chronic hypoxemia, leading to various secondary health implications, including hypertension, cardiovascular disease, and cognitive impairment.
 Continuous Positive Airway Pressure (CPAP) has traditionally served as the primary treatment for Obstructive Sleep Apnea (OSA). However, challenges with patient tolerance often hinder satisfactory compliance with this device. An alternative therapy is Inspire, an FDA-approved hypoglossal nerve stimulator. The hypoglossal nerve controls tongue protrusion and retraction. By stimulating specific branches that induce forward tongue protrusion, the muscles in the neck become rigid, preventing collapse and subsequent airway obstruction. This stimulation is synchronized with the inhalation phase, ensuring that the tongue protrudes during inspiration — the period when neck muscles are most susceptible to collapse and airway obstruction.

\begin{figure}
  \centering
  \includegraphics[width=0.43\columnwidth]{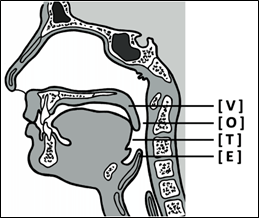}
  \includegraphics[width=0.53\columnwidth]{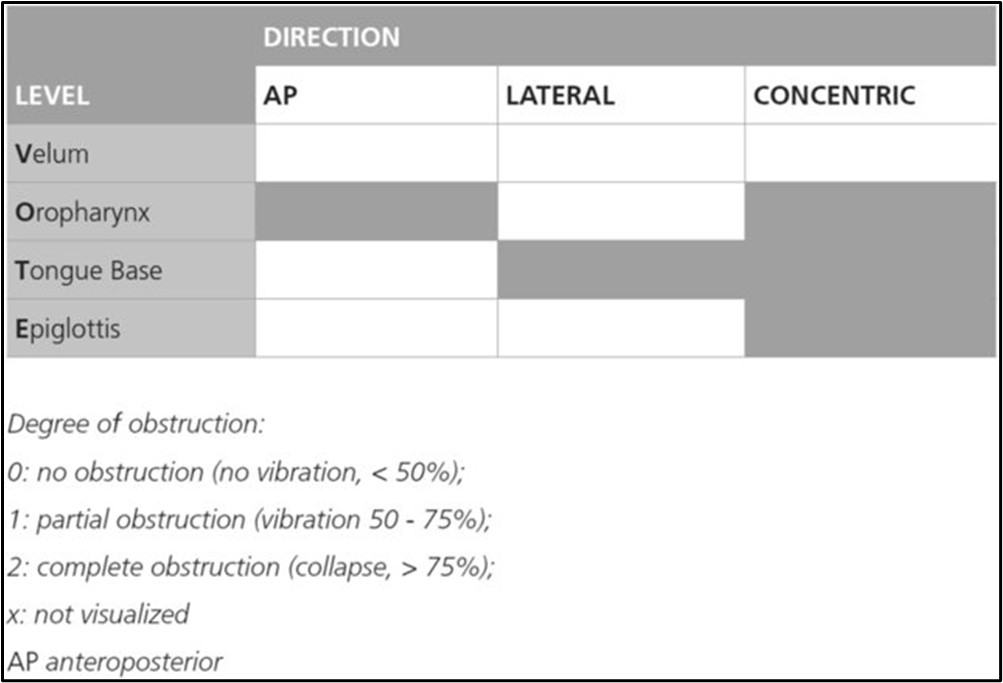}
  \caption{The VOTE score criteria and classification \cite{kastoer2018comparison}\cite{qian2016classification}.}
  \label{fig:vote}
\end{figure}

The evaluation of eligibility for Inspire involves a procedure called Drug-Induced Sleep Endoscopy (DISE). This entails administering a light amount of anesthesia to simulate sleep. The airway is then observed at four distinct locations: the velopharynx, oropharynx, tongue base, and epiglottis (VOTE). At each location, the surgeon assesses the orientation and degree of collapsibility of the airway using the VOTE score. This score is crucial in determining the most suitable therapies for patients, as specific patterns of airway collapse may not respond uniformly to all treatments. Fig. \ref{fig:vote} illustrates the location of the velopharynx, oropharynx, tongue base, and epiglottis in the airway and the VOTE score criteria and classification \cite{kastoer2018comparison}\cite{qian2016classification}.

For example, the Inspire Hypoglossal Nerve stimulator is approved for patients with predominant anterior-posterior velopharyngeal collapse. This means that this therapy is mostly beneficial for patients with anterior and posterior wall collapse during inspiration at the level of the velopharynx. If patients had concentric collapse, meaning the whole airway shrinks in the shape of a circle, at the level of the velopharynx, the Inspire device is less likely to alleviate airway obstruction. Similarly, if patients had predominant lateral wall collapse, meaning the sides of the upper airway collapse inward, the Inspire device is again less likely to be therapeutic. Additional FDA criteria for the hypoglossal nerve stimulator include $AHI \geq 15$, $AHI \leq 100$, and a $BMI \leq 40 kg/m^2$.


\section{Dataset}

In this study, we collected data from 127 patients diagnosed at the Department of Otolaryngology-Head and Neck Surgery at the University of Kansas Medical Center (KUMC). The endoscopic images were captured during DISE, a procedure involving patient sedation to simulate sleep while evaluating the upper airway for eligibility for surgical treatment for OSA. The dataset comprises 24,777 images of throat regions focusing on the base of the tongue (BOT) or velopharynx (VP). This dataset is categorized into ``responder" and ``non-responder" classes, indicating the respective patient groups that exhibited a ``response to the therapy" and those who ``did not respond to the therapy," as determined by Sher Criteria. (Sher Criteria – \(AHI<15\) events/hour and \(>50\%\) decrease in post-operative AHI from pre-op). The characteristics of the dataset are shown in Table \ref{tab:charDB}. 

\begin{figure}[t]
  \centering
  \includegraphics[width=1\columnwidth]{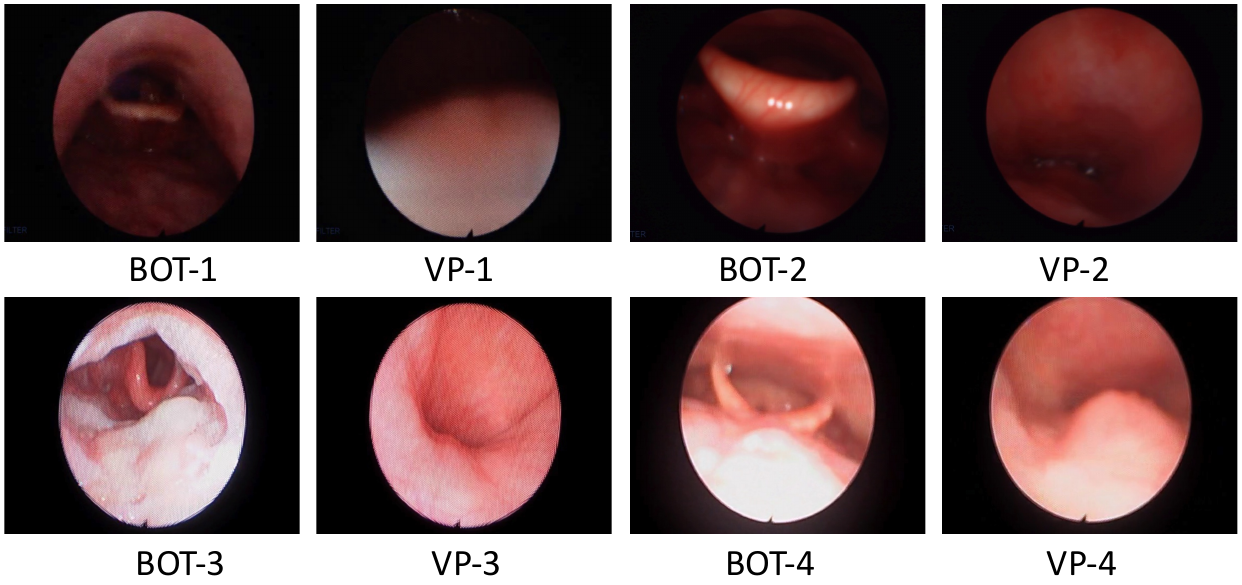}
  \caption{Sample image frames of four patients sampled from their corresponding BOT and VP sequences. Two patients are responders (1st row) and two are non-responders (2nd row).}
  \label{fig:sample}
\end{figure}

Fig. \ref{fig:sample} displays sample images of four patients, two categorized as ``responder" and two as ``non-responder," showcasing both BOT and VP image frames for each patient. Notably, distinguishing between these two patient classes based solely on their BOT or VP images proves challenging, even for experienced otolaryngologists and clinical experts in this field. The imaging process introduces significant variations in illumination, viewpoint, occlusion, and reflection, as evident in these images. Moreover, images from the same patient may exhibit substantial visual differences, further complicating the problem. One objective of this study is to investigate whether current deep learning techniques can autonomously extract pertinent features capable of discerning between ``responder" and ``non-responder" based on their BOT and/or VP images.

\begin{table*}[t]
  \caption{Characteristics of the dataset}
  \label{tab:charDB}
  \begin{tabular}{ccccccc}
    \toprule
    Total Patients & \multicolumn{2}{c}{Patients/Class} &\multicolumn{2}{c}{Images/Class} &\multicolumn{2}{c}{Frame Resolution}\\
    \midrule
    &Responder & Non-responder & Responder & Non-responder & Min & Max\\\cmidrule{2-7}%
    127& 88 & 39 & 16,515 & 8,262 & 720$\times$480 & 1920$\times$1080\\
    \bottomrule
\end{tabular}
\end{table*}
We collected two video sequences for each patient corresponding to BOT and VP. Thus, we created two distinct datasets for binary classification to assess prediction accuracy based on different throat regions. The VP dataset comprises a total of 11,298 images, with 64\% belonging to responders and 36\% to non-responders. The BOT dataset consists of 13,479 video images with approximately 69\% responders and 31\% non-responders. To obtain more statistically meaningful results, we employ 10-fold cross-validation by randomly dividing the entire dataset into 10 equal patient-wise folds. Due to variations in the number of collected frames from different patients, our training and test datasets exhibit diverse ranges. Additionally, we compare the results obtained from these two datasets in combination. The dataset sizes used for all deep learning classification models are detailed in Table \ref{tab:DBsize}.

\begin{table}
  \caption{Statistical information of the dataset.}
  \label{tab:DBsize}
    \begin{tabular}{cccc}
    \toprule
    Datasets & Images & \multicolumn{2}{c}{Images/Class} \\
    \midrule
    &&Responder & Non-responder  \\
    \midrule
    VP& 11,298 & 7,257 & 4,041  \\
    \midrule
    BOT& 13,479 & 9,258 & 4,221  \\
    \midrule
    Combined (VP+BOT)& 24,777 & 16,515 & 8,262 \\
    \bottomrule
\end{tabular}
\end{table}

In addition to the video datasets, we gathered comprehensive clinical information for all patients. The clinical data encompasses 22 essential features, including race, ethnicity, BMI level, Pre-operative Apnea-Hypopnea Index, Sleep Apnea Severity, OSA-severity, responder status, and more. We explored the prospect of making predictions solely based on the clinical data. To achieve this, we implemented and compared five machine learning-based classification models. We applied 10-fold cross-validation using the same split as the video data for a robust evaluation of model performance.

\section{Methods}

In this paper, we implemented and benchmarked the performance of six deep learning models on the video datasets and five ML algorithms on the clinical dataset. In recent years, convolutional neural networks (CNN) have achieved huge success in image classification \cite{mcclannahan2021classification}\cite{zhang2023gender}, object detection \cite{bur2023interpretable}\cite{li2021sgnet}, and segmentation \cite{patel2022fuzzynet}\cite{xiao2023edge}. We implemented the following classical CNN models.

\begin{itemize}
\item \textbf{VGG} is known for its simplicity and effectiveness. It won 1st and 2nd place in object detection and classification in 2014 ImageNet Large-Scale Visual Recognition Challenge (ILSVRC). \cite{simonyan2014very}.
\item \textbf{ResNet} was introduced to address the vanishing gradient problem in deep networks via residual blocks, allowing for the training of very deep networks \cite{jian2016deep}. Two popular structures are ResNet-50 and ResNet-101.
\item \textbf{EfficientNet-B0} was designed to achieve better performance with fewer parameters by introducing compound scaling to balance network depth, width, and resolution \cite{tan2019efficientnet}.
\item \textbf{DenseNet} introduced densely connected blocks, where each layer receives input from all preceding layers, promoting feature reuse and efficient parameter utilization \cite{huang2017densely}. DenseNet-121 and DenseNet-169 are two popular structures.
\end{itemize}


The following five ML algorithms have been implemented.
\begin{itemize}
\item \textbf{Logistic Regression (LR)} is commonly used for binary classification. The model is well-suited for linearly separable problems, but may struggle with complex relationships \cite{sperandei2014understanding}.
\item \textbf{Decision Trees (DT)} utilizes interpretable tree structure for hierarchical decision-making. It is simple but prone to overfitting on complex datasets \cite{ke2017lightgbm}.
\item \textbf{Random Forest (RF)} is an ensemble method that constructs multiple decision trees during training and outputs the mode of the classes. It can handle complex relationships but may lack interpretability \cite{biau2016random}.
\item \textbf{k-Nearest Neighbors (k-NN)} makes predictions based on the majority class or average of the k-nearest data points in the feature space. It is computationally expensive for large datasets \cite{peterson2009k}.
\item \textbf{Gradient Boosting (GB)} is an ensemble learning technique that builds a series of weak learners and combines their predictions to form a strong learner. It is powerful for classification but is computationally intensive \cite{ke2017lightgbm}.
\end{itemize}

\begin{table*}
  \caption{The implementation specifics and hyperparameter settings for each model. }
  \label{tab:settings}
  \begin{tabular}{lllll}
    \toprule
    Key settings & VGG-16 &ResNet-50/101 &EfficientNet &DenseNet-121/169\\
    \midrule
    Batch size&32&64&32&64\\
    Learning rate &0.0001&0.0001&0.015&0.00001\\
    Loss function & NLLLoss & Cross-entropy & NLLLoss &Cross-entropy\\
    Optimizer &ADAM& SGD& ADAM &ADAM\\
    Dropout &0.4&-&-&0.5\\
    Epochs &50 &50 &50 &50\\
\bottomrule
\end{tabular}
\end{table*}

\textbf{Evaluation metrics.}
To assess the performance of our models, we employed several standard evaluation metrics in image classification, including precision, recall, F1 score, AUC score, and overall accuracy \cite{vyas2024predicting}. These metrics serve as benchmarks to gauge the effectiveness of our models.
Accuracy quantifies the overall correctness of the model's predictions. It is calculated as the ratio of correctly classified instances to the total instances. 
Precision represents the ratio of true positive predictions to the total predicted positives, offering insights into the model's ability to minimize false positives, particularly when their associated costs are high. 

Recall measures the ratio of true positive predictions to the total actual positives, reflecting the model's proficiency in identifying all relevant instances, which is crucial when the cost of false negatives is significant.
The F1 score, a harmonic mean of precision and recall, provides a balanced assessment of the model's performance, particularly valuable in scenarios with class imbalances.
The AUC (area under the ROC curve) score is a metric tailored for evaluating classification models, especially in binary classification tasks. The ROC (receiver operating characteristic) curve visually represents the trade-off between sensitivity (true positive rate) and specificity (true negative rate) at various thresholds, with the AUC score quantifying the area under this curve.

\section{Experiments}
\subsection{Evaluation of Deep Learning Models}

\textbf{Pre-processing.}
In this study, the original dataset spans a diverse range of resolutions, ranging from $720\times480$ to $1920\times1080$. However, the developed deep neural networks require input images of a fixed size of $224\times224$. To align with this requirement, we initially downscaled all images in both BOT and VP datasets to the expected input size of the networks. During the data augmentation phase, we introduced random horizontal and vertical shifts within a range of 0.5 and applied rotations up to 35 degrees. 

\textbf{Training settings.}
We conducted an evaluation of six deep learning networks, i.e., VGG-16, ResNet-50, ResNet-101, EfficientNet-B0, DenseNet-121, and DenseNet-169, utilizing the two DISE medical image datasets. Additionally, a comparison was made with a combined dataset. Implementation specifics and hyperparameter settings for each model are outlined in Table \ref{tab:settings}. All models undergo pre-training on the ImageNet dataset with the top layer fine-tuned using our training set, with a 40\% dropout for VGG-16 and a 50\% dropout for DenseNet-121/169. All models were trained for 50 epochs. All experiments were executed in PyTorch using NVIDIA A100 GPU.

\begin{table}
  \caption{Patient level accuracy on VP and BOT datasets}
  \label{tab:patient_acc}
  \begin{tabular}{ccc}
    \toprule
    Method&Patient acc (VP)&Patient acc (BOT)\\
    \midrule
    VGG-16&0.711$\pm$0.04&0.642$\pm$0.05\\
    ResNet-50&0.636$\pm$0.05&\textbf{0.697}$\pm$0.15\\
    ResNet-101&0.631$\pm$0.03&0.693$\pm$0.11\\
    DenseNet-121&0.659$\pm$0.09&0.645$\pm$0.14\\
    DenseNet-169&\textbf{0.713}$\pm$0.08&0.626$\pm$0.12\\
    Efficient-B0&0.505$\pm$0.08&0.594$\pm$0.08\\
\bottomrule
\end{tabular}
\end{table}

\textbf{Model performance.}
Given the limited sample size of the collected dataset, all models underwent evaluation using 10-fold cross-validation to yield less biased results. Consequently, the data were randomly divided into 10 equal groups at the patient level to prevent cross-contamination of the training and test sets. Thus, we conducted 10 experiments for each model, alternating the use of one subset as the test data while training the model with the remaining data. The statistical performance of all models is detailed in Table \ref{tab:DL}, presenting mean values for training accuracy, validation accuracy, F1 score, and AUC score for each model. Insights drawn from the results include: (i) All models exhibited higher accuracy on the Velopharynx (VP) dataset compared to BOT and the combined datasets, indicating that the velopharynx (VP) area contains more discriminative features for classification than the base of the tongue (BOT). (ii) Most models consistently performed well in our experiments. In contrast, DenseNet-169 achieved the highest accuracy, while EfficientNet yielded lower performance compared to other models.
\begin{figure*}[t]
  \centering
  \includegraphics[width=0.85\textwidth]{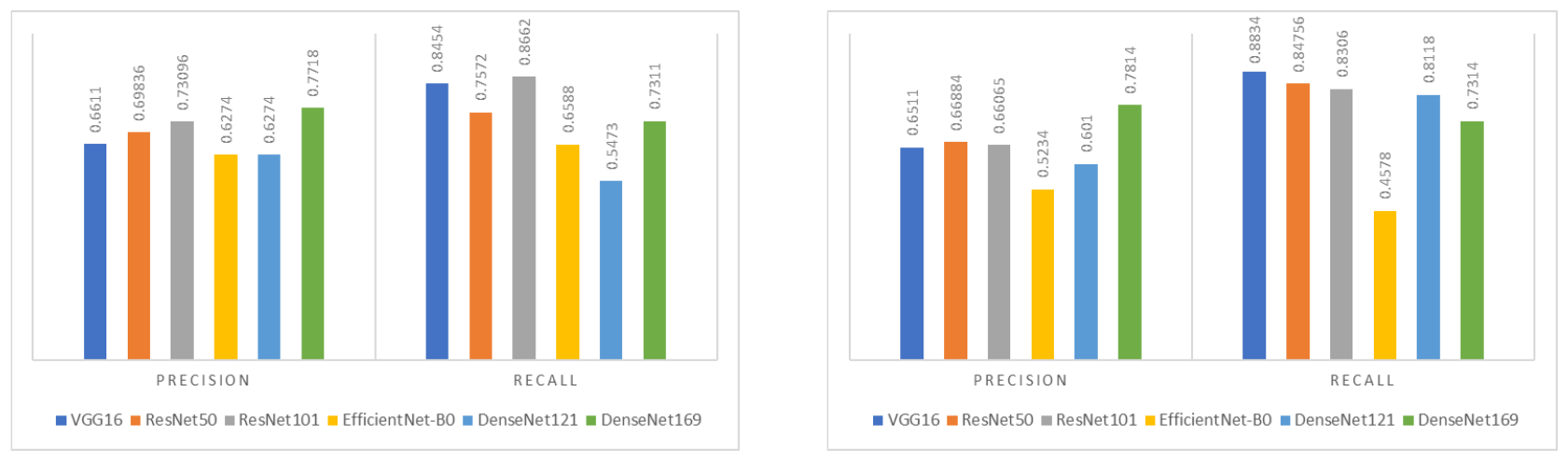}
  \caption{The precision and recall of all models on BOT (left) and VP (right) datasets.}
  \label{fig:PR}
\end{figure*}

\begin{table*}
  \caption{Performance of deep learning models (mean$\pm$std).}
  \label{tab:DL}
  \begin{tabular}{lcccc}
    \toprule
    Method\_DB & Frame acc & Train acc & F1 score & AUC\\
    \midrule
    VGG-16\_VP&0.676$\pm$0.04&\textbf{0.833}$\pm$0.05&0.780&\textbf{0.595}\\
    VGG-16\_BOT&0.608$\pm$0.05&0.799$\pm$0.03&0.725&0.512\\
    VGG-16\_combine&0.636$\pm$0.04&0.780$\pm$0.04&0.763&0.494\\
    \midrule
    ResNet-50\_VP&0.626$\pm$0.03&0.748$\pm$0.02&0.745&0.519\\
    ResNet-50\_BOT&0.608$\pm$0.13&0.755$\pm$0.04&0.729&0.504\\
    ResNet-50\_combine&0.642$\pm$0.05&0.734$\pm$0.02&0.765&0.501\\
    \midrule
    ResNet-101\_VP&0.626$\pm$0.03&0.722$\pm$0.01&0.732&0.509\\
    ResNet-101\_BOT&0.675$\pm$0.11&0.731$\pm$0.02&\textbf{0.790}&0.501\\
    ResNet-101\_combine&0.674$\pm$0.05&0.726$\pm$0.01&0.711&0.497\\
    \midrule
    DenseNet-121\_VP&0.633$\pm$0.08&0.797$\pm$0.02&0.711&0.464\\
    DenseNet-121\_BOT&0.635$\pm$0.15&0.708$\pm$0.04&0.680&0.464\\
    DenseNet-121\_combine&0.621$\pm$0.04&0.784$\pm$0.05&0.682&0.461\\
    \midrule
    DenseNet-169\_VP&\textbf{0.691}$\pm$ 0.09&0.712$\pm$0.03&0.728&0.510\\
    DenseNet-169\_BOT&0.642$\pm$0.13&0.721$\pm$0.04&0.748&0.509\\
    DenseNet-169\_combine&0.682$\pm$0.06&0.778$\pm$0.06&0.712&0.504\\
    \midrule  
    EficientNet\_VP&0.522$\pm$0.08&0.539$\pm$0.03&0.400&0.500\\
    EfficientNet\_BOT&0.606$\pm$0.09&0.565$\pm$0.04&0.578&0.500\\
    EfficientNet\_combine&0.506$\pm$0.08&0.562$\pm$0.02&0.445&0.500\\
\bottomrule
\end{tabular}
\end{table*}

\textbf
All deep learning models in this study provide predictions for individual images, and accuracy is computed at the image level. However, since the medical dataset was categorized into responder and non-responder classes based on patients, it is more meaningful to make predictions at the patient level (i.e., sequence level). In this paper, we introduce patient accuracy, which signifies accuracy at the patient level by calculating the majority of the frame-level accuracy within each sequence. As depicted in Table \ref{tab:patient_acc}, the accuracy is significantly enhanced for most cases at the patient level. For instance, the accuracy for VGG corresponding to VP, BOT, and combined datasets increases from 67.6\%, 60.8\%, and 63.6\% to 71.1\%, 64.2\%, and 68.9\%, respectively. Overall, both VGG and DenseNet-169 demonstrate the best performance for the VP dataset.

\textbf{Precision and Recall.}
Precision and recall serve as critical metrics in classification tasks, offering insights beyond mere accuracy, particularly valuable for imbalanced datasets or scenarios where the cost of false positives and false negatives varies significantly. In Fig. \ref{fig:PR}, the precision and recall of all DL models on BOT and VP datasets are illustrated. Notably, DensNet-169 achieved the highest precision rates of 78.14\% for VP and 77.18\% for BOT. Meanwhile, VGG-16 and ResNet-101 showcased competitive high recall values across the two datasets. Precision quantifies the accuracy of positive predictions by the classifier, while recall measures the classifier's proficiency in identifying all relevant instances of the positive class. Striking a balance between precision and recall often involves trade-offs tailored to specific needs in practice.


\subsection{Evaluation of Machine Learning Models}

\textbf{Training settings.}
All machine learning models were assessed using the clinical information dataset. The evaluation employed a 10-fold cross-validation approach with identical data splits as those used for the video datasets. Consequently, the training and test sets in each trial shared the same patient IDs as those utilized in the training of deep neural network models. We leverage the Scikit-learn library \cite{pedregosa2011scikit} to implement the ML models, and the hyperparameters and settings for these models are detailed below.
\begin{itemize}
    \item \textbf{Decision Tree:} Criterion = `gini'; Max Depth = None; Min Samples Leaf = 10; Min Samples Split = 2 
    \item  \textbf{Logistic Regression:} C = 0.001; Penalty = `none'; Solver = `sag'
    \item \textbf{Gradient Boosting}: Learning Rate = 0.01; Max Depth = 5; Max Features = `log2'; N Estimators = 200; Subsample = 0.7
    \item \textbf{Random Forest:} Bootstrap = True; Max Depth = 5; Min Samples Leaf = 1; Min Samples Split = 8; N Estimators = 100
    \item \textbf{k-Nearest Neighbors:} Metric = `euclidean'; N Neighbors = 30; Weights = `uniform'
\end{itemize}

\textbf{Model performance.}
The performance of all ML models is assessed using the following metrics: F1 score, AUC score, and overall accuracy. The evaluation results are presented in Table \ref{tab:ML}, from which we can see that the performance of all ML models closely aligns with that of the DL models. Among all ML models, Logistic Regression stands out with the highest accuracy at 68.5\% and the highest F1 score at 0.804, closely followed by k-Nearest Neighbors and Gradient Boosting. A higher F1 score indicates a better balance between precision and recall, positioning Logistic Regression as the top-performing model in terms of accuracy and F1 score. It is noteworthy that, although Gradient Boosting achieves an accuracy of 64.2\%, slightly lower than LR and k-NN, it attains the highest AUC score of 0.531 among all models. AUC assesses the model's ability to distinguish between positive and negative instances.
    
\begin{table}[t]
  \caption{Performance of machine learning models.}
  \label{tab:ML}
  \begin{tabular}{lccc}
    \toprule
    Algorithm & Accuracy & F1 Score & AUC\\
    \midrule
    Decision Tree&$0.578\pm0.10$&$0.692$&$0.506$\\
    Gradient Boosting&$0.642\pm0.14$&$0.749$&$\textbf{0.531}$\\
    k-Nearest Neighbors&$0.675\pm0.06$&$0.804$&$0.493$\\
    Logistic Regression&$\textbf{0.685}\pm0.02$&$\textbf{0.813}$&$0.494$\\
    Random Forest&$0.636\pm0.13$&$0.746$&$0.519$\\
\bottomrule
\end{tabular}
\end{table}

In this study, the leading ML algorithm and the top DL model demonstrated comparable accuracy. DL models have the ability to automatically learn hierarchical representations of features from raw data, particularly at capturing intricate patterns in extensive and complex datasets, thereby eliminating the need for manual feature engineering. However, DL models are computationally expensive and demand substantial training data. On the other hand, classical ML algorithms exhibit less dependence on data and computational efficiency. Nevertheless, they necessitate manual feature engineering, relying on domain knowledge for feature collection and extraction that may be challenging for medical applications where relevant information is often limited. A promising approach is to leverage the strengths of both techniques by combining them, aiming to capitalize on the advantages each offers.

\section{Conclusion}
In this paper, we have showcased the potential of employing machine learning and deep learning models to forecast patient responsiveness to Inspire therapy. Two datasets were meticulously collected and annotated from Drug-Induced Sleep Endoscopy (DISE) videos, alongside a clinical information dataset of 127 patients. The performance of six deep learning models and five machine learning models was implemented and evaluated using the curated datasets. The insights derived from this study serve as a valuable reference for future research in this domain. Acknowledging the limitations imposed by the datasets' restricted size and recognizing the confined accuracy achieved when using individual datasets, we are presently developing a multimodal fusion framework. This framework aims to enhance the performance of existing learning models by exploring both video and text data.




\begin{acks}
The project is partly supported by the Natural Sciences and Engineering Research Council of Canada (NSERC) and the National Institutes of Health (NIH).
\end{acks}

\balance
\bibliographystyle{ACM-Reference-Format}
\bibliography{References}

\appendix

\end{document}